\documentclass[mathleft
]{an}
\usepackage{graphicx}
\usepackage{times}
\usepackage{epsfig}

\def\xmm{{\sl XMM-Newton}}

\def\gtrsim{\lower 2pt \hbox{$\, \buildrel {\scriptstyle >}\over
{\scriptstyle \sim}\,$}}
\def\lesssim{\lower 2pt \hbox{$\, \buildrel {\scriptstyle <}\over
{\scriptstyle \sim}\,$}}

\def\ovii{O~{\scriptsize VII}}
\def\oviii{O~{\scriptsize VIII}}

\def\neix{Ne~{\scriptsize IX}}
\def\nex{Ne~{\scriptsize X}}
\def\mgxi{Mg~{\scriptsize XI}}

\def\HII{H~{\scriptsize II}}
\def\fexvii{Fe~{\scriptsize XVII}}
\begin{document}

\Pagespan{789}{}
\Yearpublication{2006}%
\Yearsubmission{2005}%
\Month{11}%
\Volume{999}%
\Issue{88}%

\title{Spectroscopic Evidence of Charge Exchange X-ray Emission from Galaxies}

\author{Q. Daniel Wang\inst{1}\fnmsep\thanks{
  \email{wqd@astro.umass.edu}\newline}
\and  Jiren Liu\inst{2}
}
\institute{
University of Massachusetts, Amherst, USA
\and 
National Astronomical Observatories, Chinese Academy of Sciences, Beijing, China
}

\received{1 Feb 2012}
\accepted{}


\abstract{%
What are the origins of the soft X-ray line emission from non-AGN galaxies? 
{\sl XMM-Newton} RGS spectra of nearby non-AGN galaxies (including starforming ones:
M82, NGC 253, M51, M83, M61, NGC 4631, M94, NGC 2903, and the Antennae galaxies, 
as well as the 
inner bulge of M31) have been analyzed. In particular,  the K$\alpha$ triplet of 
\ovii\  shows that the resonance line is typically weaker than the forbidden and/or 
inter-combination lines. This suggests that a substantial fraction of the 
emission may not arise directly from optically thin thermal plasma, as commonly
assumed, and may instead originate at its interface with neutral gas
via charge exchange. This latter origin naturally explains the observed spatial correlation of 
the emission with various tracers of cool gas in some of the galaxies.
However, alternative scenarios, such as the resonance scattering by the
plasma and the relic photo-ionization by AGNs in the recent past,
cannot be ruled out, at least in some cases, and are being examined. 
Such X-ray spectroscopic studies are important to the understanding of
the relationship of the emission to various
high-energy feedback processes in galaxies.  }
\maketitle

\section{Introduction}
Soft X-ray line emission has commonly been used to trace various 
types of galactic feedback in nearby starburst and normal galaxies. Assuming an origin of this
emission in optically thin thermal (collisionally excited) hot plasma, one may further estimate 
its mass, energy, and chemical contents and even its outflow rate from such a galaxy. 
However, the X-ray emission can be seriously contaminated, if not dominated, by 
various other contributions. They must be quantified before we can reliably use 
the emission to study the galactic feedback and its impact on the 
galactic ecosystem.

Here we focus on the evidence of a significant contribution from 
charge exchange  (CX, also called charge transfer) X-ray emission 
at interfaces between hot and cool gases in galaxies. In the simplest 
form, a CX can be represented by
\begin{equation}
{\rm A^{q+} + N \rightarrow A^{(q-1)+*} + N^{+}},
\end{equation}
where a highly ionized ion ${\rm A^{q+}}$ (like \oviii\ or \nex) picks up an electron 
from a neutral species N (like H, ${\rm H_2}$, or He), 
producing an excited ion ${\rm A^{(q-1)+*}}$, 
which then decays to the ground state, emitting X-ray photon(s). 
Extensive studies have been carried out on the solar wind CX 
(see Dennerl 2010 for a recent review). In particular, the solar wind
CX origin of the X-ray emission from comets has been established spectroscopically. 
The solar wind CX with the local interstellar medium may also
account for a substantial part of the observed soft
X-ray background (e.g., Koutroumpa et al. 2011). Furthermore,
the CX has been speculated to be important in many other
types of astrophysical objects 
(e.g., Lallement 2004) and has specifically been proposed to be responsible for
emission features unaccounted for by thermal plasma modeling alone in X-ray CCD
spectra of Galactic supernova remnants (e.g., Katsuda et al. 2011) and giant \HII\ regions (e.g., Townsley et
al. 2011). Such CX scenarios, if confirmed spectroscopically, 
could provide new insights into the physical, chemical, and kinematic properties of
the underlying hot plasma and its interplay with cool gas. 

The most straight-forward X-ray spectroscopic diagnostic of a significant
CX contribution is the so-called G ratio [$=(f +i)/r$, forbidden + inter-combination to resonance lines]
of the K$\alpha$ triplets of He-like ions. Take \ovii\ as an example.
For an optically thin thermal plasma with a temperature of a few $10^6$ K
and in collisional ionization equilibrium, G should be considerably less than one
(Porquet et al. 2001). The temperature may be
constrained independently, for example, by the resonance transition
ratios of H-like to He-like ions (e.g., Liu et al. 2011). 

The only available instruments that allow for such X-ray spectroscopy of moderately 
extended X-ray-emitting regions are the 
\xmm\ Reflection Grating Spectrometers (RGSs), which have both a large 
effective area and a large dispersion power.
\xmm\ observations, with a spatial resolution of $\sim
6^{\prime\prime}$ (FWHM) at $\sim 1$ keV, are typically
carried out with simultaneous exposures by the two co-aligned RGSs, together with
the European Photon Imaging Cameras (X-ray CCDs).
The orientation of the satellite determines the dispersion direction of the
RGSs, which are slit-less spectrometers and are sensitive to photons 
in the 0.3-2 keV range. For a point-like source, the spectral resolution of 
the RGSs is $\lambda/\delta\lambda \sim 400$ (HEW) 
at $\sim 20$ \AA. For an extended region, one can obtain useful 1-D spatial
information over the cross-dispersion range of 5$^\prime$ wide. In the
dispersion direction, the RGSs cover the entire 
field of view of the telescope mirrors. The extent of the region translates to
an apparent (1st-order) wavelength shift 
of 0.14 \AA~arcmin$^{-1}$ with respect to a non-displaced
source. Therefore, for a region with a moderate extent (e.g., $\lesssim 2^\prime$),
the spectral resolution of an
integrated RGS spectrum can still be
substantially better than that of X-ray CCD data.
In fact, the wavelength shift can sometimes be used 
to extract the spatial information along the dispersion direction. 

\section{Examples}

\subsection{M31 Bulge}

\begin{figure*}[!tbh]
  \centerline{
  \epsfig{figure=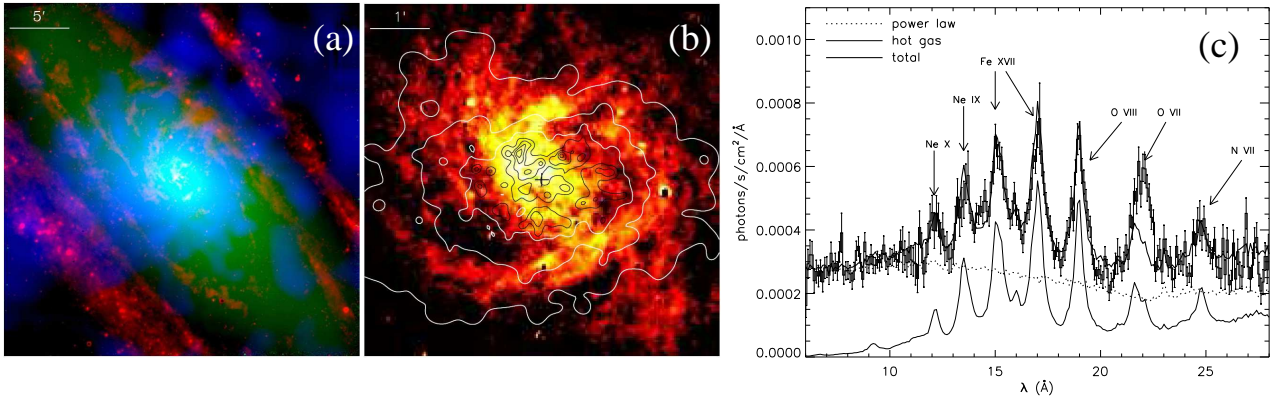,width=1\textwidth}
  }
  \caption{\footnotesize  
{\bf (a)} Tri-color image of the spheroid region of M31: 
{\sl Spitzer}/MIPS 24 $\mu$m emission ({\sl red}),
   2MASS K-band emission ({\sl green}), and
   {\sl Chandra} 0.5-2 keV emission of truly diffuse hot gas ({\sl blue}).
The data like these show clear evidence for a subsonic outflow of hot gas
(e.g., Li \& Wang 2007; Li et al. 1011; Wang 2010 and references therein).
    {\bf (b)} Intensity contours of the  diffuse soft X-ray emission overlaid
    on an H$_{\alpha}$ image of the M31 central region. The contours are at 5, 10, 19, 27, 35, 45 and
55 $\times 10^{-3} {\rm~cts~s^{-1}~arcmin^{-2}}$. 
The plus sign marks the M31. A detailed 
analysis of the data shows evidence for strong interaction between the hot 
and cool gas phases, which may provide a mechanism for starving the nucleus
as marked by the plus sign (Li et al. 2009). 
{\bf (c)} Fluxed \xmm\ RGS spectrum of the M31 central region (Liu et al. 2010). Both RGS1 and RGS2 spectra are combined together. The curves 
show the fit with a single-temperature thermal plasma, plus a power-law that characterizes 
the discrete source contribution. Notice that the centroid of the \ovii\ line
is closer to $f$  wavelength than the $r$ wavelength. [Figures are 
reproduced from Li et al. (2009) for (a) and (b) and from Liu et al.
(2010) for (c).]}
\label{fig:m31}
\end{figure*}

Our existing study of the central bulge of M31 (e.g., Fig.~\ref{fig:m31}) 
demonstrates the spectroscopic diagnostic power
of RGS observations in probing the nature of the soft X-ray line emission 
(Liu et al. 2010). Fig.~\ref{fig:m31}c, for example, shows
a clear intensity excess of the observed spectrum above a simple thermal 
plasma model fit at $\sim 22$ \AA. This excess is apparently due to the 
prominence of the forbidden line 
(at 22.1 \AA) with respect to the resonant one (21.6 \AA) of the He-like 
\ovii\ K$\alpha$ triplet. Furthermore,
the excess is shown to be present throughout much of the cross-dispersion range
of the RGS observations and seems to correlate well with the 
distribution of the cool gas, as traced by the H$\alpha$ emission 
(Fig.~\ref{fig:m31}b; Liu et al. 2010). 
In addition, the centroid of 
the  \neix\ K$\alpha$ triplet is also closer to its forbidden line wavelength
than to its resonant one (Fig.~\ref{fig:m31}c).
A natural explanation of these results is that the excess represents
the contribution from the CX between the two gas phases. The presence of
the diffuse hot gas in the bulge is expected from the stellar feedback
in forms of mass loss and Type Ia SNe of low-mass stars (e.g., 
Tang et al. 2010). There is little evidence
for any recent massive star formation in the bulge or significant AGN 
activity  (Li et al. 2009, 2011). The CX and other heating/evaporation 
processes at cool/hot gas interfaces may play an important role in launching
a bulge-wide, mass-load outflow and in starving the supermassive black hole at
the center of the galaxy (Li \& Wang 2007; Tang et al. 2009). Similar processes
may also happen in other galactic spheroids.

\subsection{Active Star-forming Galaxies}

\begin{figure*}[!tbh]
\centering
\epsfig{figure=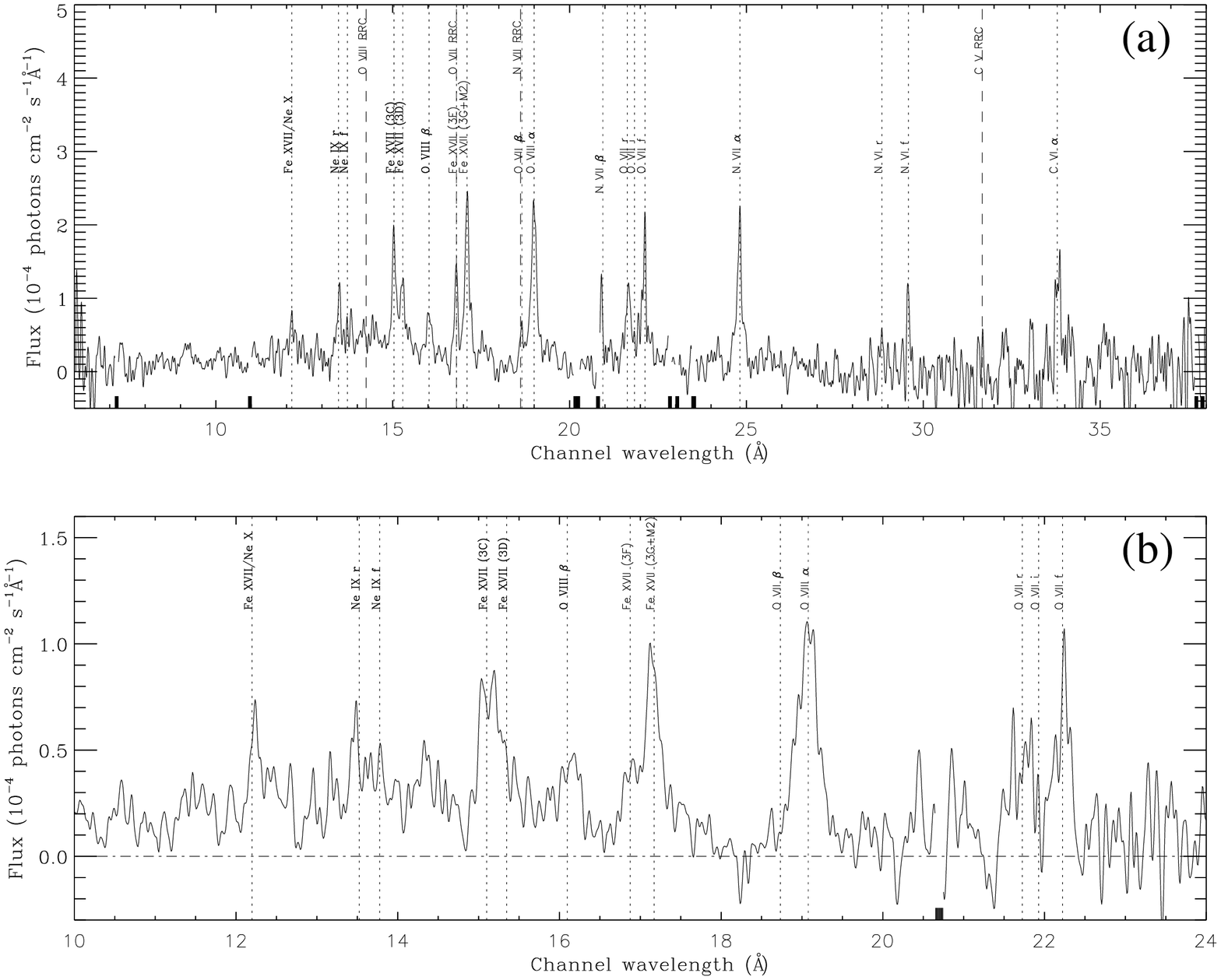,width=1.0\textwidth}
 \caption{ \footnotesize 
Fluxed RGS spectra of the central region of M51 (a; with a total effective 
exposure of 97 ks) and the
Antennae galaxies (b; 172 ks).
Spectral bins with no exposure (e.g., due to the RGS CCD gaps) are 
marked with bars at the wavelength axis. Positions of key lines or recombination
edges are marked.}
\label{m51_spec}
\end{figure*}

\begin{figure*}[!th]
\centering
\includegraphics[height=2.3in]{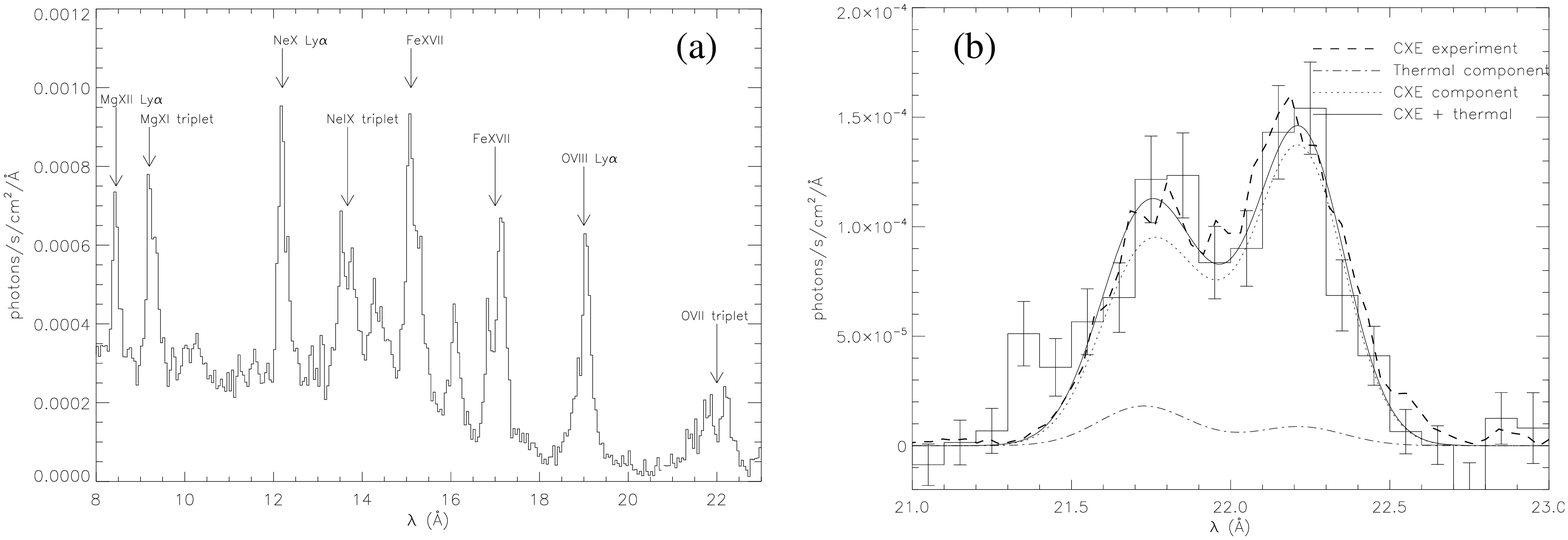}
\caption{\footnotesize (a) Fluxed RGS spectrum of M82 extracted from 
a central 1$^\prime$ region (Liu et al. 2011; see also Ranalli et al. 2008). 
The continuum is mainly due to X-ray binaries. (b) Close-up of the
\ovii\ K$\alpha$ triplet (Liu et al. 2011). 
A continuum, which is interpolated between the line wings, has been subtracted. 
[Figures are 
reproduced from Liu et al. (2011)]} 
\label{m82_specs}
\end{figure*}

RGS spectra of star-forming galaxies, as well as individual starburst regions, 
show roughly similar X-ray spectroscopic characteristics. 
Fig.~\ref{m51_spec} presents two example spectra. Our
existing analysis has focused on the \ovii\ K$\alpha$ triplet of a sample of nine nearby 
star-forming galaxies with available 
\xmm\ observations of relatively good signal to noise ratios (Liu et al. 2012). 
The observed G ratios of these galaxies are all larger than 1.2, substantially higher
than expected from the thermal emission. For NGC 253, M51, M83, M61,
NGC 4631 and the Antennae galaxies, the G ratios are consistent with the laboratory-measured ratio (2.2) of the CX, 
while for M94 and NGC 2903, a combination of thermal and CX gives better
fits. The G ratios for other He-like ions such as \neix\ (with higher ionization
energy thresholds), are typically 
much smaller and are thus more consistent with the thermal origin. 
In the central region of M82 (Fig.~\ref{m82_specs}; Liu et al. 2011), for example,
the CX contributions are 90, 50 and 30\% to the fluxes of the \ovii, \neix,
and \mgxi\ K$\alpha$ triplets, respectively; the 
averaged CX contribution is thus $\sim 50\%$ of the total observed 
triplet fluxes. The \ovii\ triplet G ratio also appears to change 
from one region to another. Furthermore, the cross-dispersion profiles of 
the \ovii\ triplet and \oviii\ Ly$\alpha$ line are 
similar, indicating that they may have a similar origin (Liu et al. 2010).

\begin{figure*}[!th]
\begin{center}
\psfig{figure=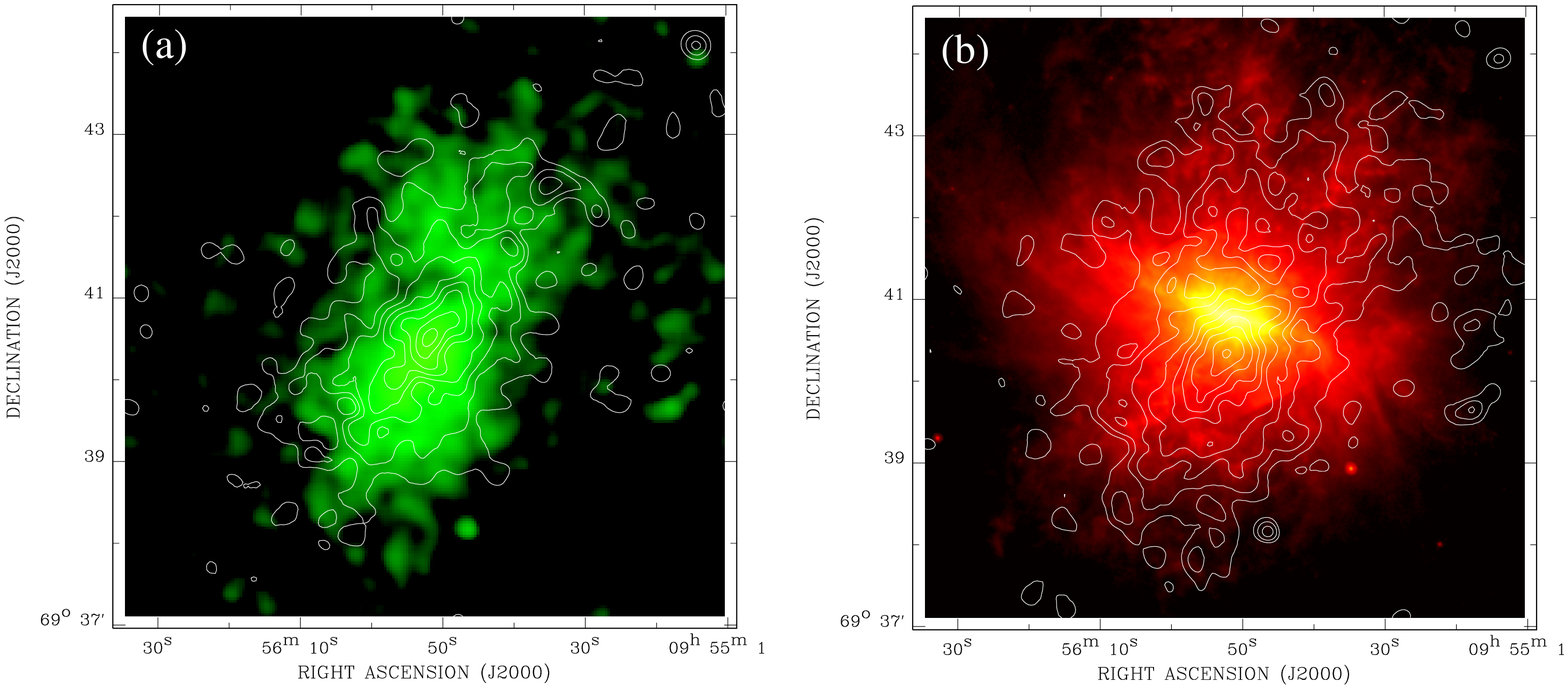,width=1\textwidth}
\end{center}
\caption{\footnotesize 
\oviii\ and \nex\ Ly$\alpha$ line intensity images of M82: (a) \nex\ contours overlaid on the \oviii\  
image; (b) \oviii\ intensity contours overlaid on the {\sl Spitzer} IRAC 8$~\mu m$ image. The 
Ly$\alpha$ line intensity distributions are reconstructed 
from five RGS observations. 
The local backgrounds (including the on-line continuum emissions chiefly from point-like 
sources) have been subtracted.
}
\label{m82_oviii_im}
\end{figure*}

\section{Implications}

Determining the origin(s) of the soft X-ray line emission is important for 
understanding the galactic feedback, hence the
evolution of galaxies. Take nuclear
starbursts as an example, which can drive so-called superwinds
and have substantial impacts on galactic ecosystem and possibly even on the
surrounding intergalactic medium. However, it remains controversial as to whether 
the observed diffuse soft X-ray emission
around starburst galaxies represents the superwinds themselves or just their interfaces 
with entrained or even in-falling cool gas clouds/filaments 
(e.g., Strickland \& Stevens 2000; Bauer et al. 2007).
Temperature and abundance anomalies have also been 
reported for starburst galaxies when diffuse X-ray emission is modeled with thermal
plasma alone. As demonstrated above, 
the modeling of the thermal emission alone is not adequate
and that other contributions or effects 
need to be accounted for in order to reliably characterize
the properties of the superwinds (e.g., Liu et al. 2011). 

If the CX is indeed significant, it will then be essential to quantify its 
contribution. In contrast to the thermal emission, the CX contributes only lines.
If an X-ray spectrum arising even partly from the CX is modeled with thermal emission
only,
then the inferred temperature and chemical properties of the plasma could  be grossly 
wrong. To measure the plasma properties more accurately, one needs to decompose
the CX and thermal contributions, spectroscopically. 
While our existing studies are based chiefly on the G ratios 
of the K$\alpha$ triplets of He-like ions (mostly \ovii;
e.g., Liu et al. 2010, 2011, 2012),
jointly fitting multiple emission lines or even an entire spectrum will 
maximize the use of the spectroscopic information available in the RGS data to 
effectively test the CX model, to tightly 
constrain multiple model parameters with proper error propagation, and to 
decompose the thermal and CX contributions. In a joint analysis, 
even spectral lines that are not individually detected can be useful, 
as demonstrated in existing X-ray absorption line spectroscopy (e.g., Yao \& Wang 2005;
Wang et al. 2010 and references therein). Such an analysis will
become possible with the upcoming CX spectral model (see the presentation by 
Smith et al. in this workshop). A joint analysis 
can, in principle, provide us with unique diagnostics of the hot plasma and its interplay with 
cool gas. The local emissivity of the CX (assuming single electron capture) is
$\sim n_{A^{q+}} n_{neu} v_r \sigma$, 
where $n_{A^{q+}}$ and $n_{neu}$ denote the ion and neutral atom densities, while
$v_r$ and $\sigma$ are their relative velocity and CX
cross-section. The $n_{A^{q+}}$ values of relevant ions (e.g., \oviii\ and \nex) 
depend on the total particle density, metal abundance, and ionization 
temperature of the hot plasma.
These parameters can thus be inferred from the emission line intensities of 
various ion species. Moreover, the relative intensities of multiple 
transitions of same ion species are also sensitive to $v_r$ (e.g.,
Greenwood et al. 2004). Therefore, the modeling of the CX contribution
can be a powerful tool in probing the thermal, chemical, and
kinematical properties of the hot plasma.

\section{Potential Complications}

While the CX appears to be an important mechanism in generating soft X-ray line
emission, we also need to consider other plausible scenarios in various 
circumstances. 
One of such scenarios is the resonance scattering of emission
line photons by the X-ray-emitting plasma itself (e.g., Porquet et al. 2001). 
Such scattering re-distributes optically thick 
line photons to outer regions.  This effect 
on the spatial distribution of the \fexvii\ $2p-3d$ transition has clearly been 
shown with the RGS observations of a few giant elliptical galaxies (Xu et al. 2002;
Werner et al. 2009). However, the effectiveness of the scattering is not 
clear in low $L_X/L_K$ spheroids (e.g., the bulge of M31), 
which contain hot plasma of relatively low column density, or in starburst galaxies,
where hot superwinds are expected to have large velocity dispersions.
Our preliminary estimates indicate that the scattering of the \ovii\ K$\alpha$ 
and \oviii\ Ly$\alpha$ resonance lines could be significant in the inner bulge region
of M31, depending largely on how small the bulk and turbulent
velocities of the plasma are. We are currently conducting detailed simulations 
to determine the significance of the 
resonance scattering, which may incidentally place constraints on the 
plasma velocities. If the resonance scattering is important, then one 
may expect to observe a systematic decreasing G ratio with increase distances 
from galaxy central regions. But this effect is not apparent
in the existing data  (Liu et al. 2012),
although a thorough analysis is yet to be performed. The effect could also be
diminished if scattered photons are largely absorbed
by cool gas. This complication also needs to be investigated. 

Large G ratios can further be produced by plasma under a non-equilibrium
ionization condition. Such a condition may be created in a cooling superwind
due to fast adiabatic expansion, for example  (e.g., Li et al. 2006). The
delayed re-combination of highly ionized ions in the plasma tends
to occur primarily in outer regions of a
starburst galaxy.  But the large G ratio is  observed in inner
regions of galaxies (Liu et al. 2010, 2012). The recombination scenario also does not explain
the good correlation of the X-ray emission with the cool gas tracers.
A similar condition, however, may also be produced by AGNs. Our sample selection of non-AGN galaxies 
should have minimized this complication (Liu et al. 2012). 
We have even avoided the inclusion of galaxies with significant Seyfert 2 nuclei, 
which could contribute to diffuse soft X-ray emission via photo-ionization and/or 
scattering. However, there might be significant contributions from AGN
relics in some galaxies. We are modeling 
the emission from diffuse plasma heated in the past, particularly as a function 
of time since the last AGN activity to check how this scenario could be spectroscopically distinguished 
from the CX. One known distinction is that a significant AGN relic 
should show the so-called recombination continuum (or edges; e.g.,
Guainazzi \& Bianchi 2007), whereas the CX should not.

In addition to the spectral analysis and modeling, spatially resolved information also needs
to be fully explored. For example, Fig.~\ref{m82_oviii_im} shows a preliminary 
version of 2-D
X-ray line emission images of M82, reconstructed from the RGS observations.
The comparison of such spatial distributions of various emission lines
will provide new tests of the scenarios.

Ultimately, we probably also need as much information as can be extracted from 
data in other multi-wavelength observations to understand the complex
high-energy phenomena and processes involved in the heating, transportation, 
and cooling of the hot plasma and in its interaction with cool gas in galaxies.

\acknowledgements
We are grateful to the organizers of the workshop for the invitation and 
hospitality and thank the referee for thoughtful comments and 
our collaborators for their contributions to some of 
the work described above. 


\begin{thebibliography}{}
  \bibitem{} Bauer M., et al. 2007, A\&A, 467, 979 
  \bibitem{} Dennerl, K. 2010, SSR, 157, 57
  \bibitem{} Greenwood, J.B., et al. 2004, PhST, 110, 358
  \bibitem{} Guainazzi, M. \& Bianchi, B. 2007, MNRAS, 374, 1290
  \bibitem{} Ji, L., Wang, Q. D., \& Kwan, J. 2006, MNRAS, 372, 497
  \bibitem{} Katsuda, S., et al. 2011, ApJ, 730, 24
  \bibitem{} Koutroumpa D., et al. 2011, ApJ, 726, 91
  \bibitem{} Lallement R., 2004, A\&A, 422, 391
  \bibitem{} Li, Z., \& Wang, Q.D.  2007, ApJL, 668, 39
  \bibitem{} Li, Z., Wang, Q.D. \& Wakker, B.~P. 2009, MNRAS, 397, 148
  \bibitem{} Li, Z., et al. 2011, ApJ, 730, 84
  \bibitem{} Liu, J., Mao, S., \& Wang, Q.D. 2011, MNRAS, 415, 64
  \bibitem{} Liu, J., Wang, Q.D., Li Z., \& Peterson, J.R. 2010, MNRAS, 404, 1879 
  \bibitem{} Liu, J., Wang, Q.D. Mao, S.  2012, MNRAS, in press
  \bibitem{} Porquet, D., et al. 2001, A\&A, 376, 1113
  \bibitem{} Ranalli, P., et al. 2008, MNRAS, 386, 1464
  \bibitem{} Strickland, D.K., \& Stevens, I.R., 2000,  MNRAS, 314, 511
  \bibitem{} Tang, S., Wang, Q. D., et al. 2009, MNRAS, 392, 77
  \bibitem{} Tang, S., \& Wang, Q. D. 2010,  MNRAS, 408, 1011
  \bibitem{} Townsley, L.K., et al. 2011, ApJS, 194, 16
  \bibitem{} Wang, Q.D. 2010, PNAS, 107, 7168
  \bibitem{} Werner, N., et al. 2009, MNRAS, 398, 23
  \bibitem{} Xu, H., et al. 2002, ApJ, 579, 600
  \bibitem{} Yao, Y. \& Wang, Q.D. 2005, ApJ, 624, 751
\end{thebibliography}
\end{document}